\documentclass[12pt]{article}
\usepackage{a4wide}
\usepackage[dvips]{graphicx}
\usepackage{amssymb}

\newcommand\be{\begin{equation}}
\newcommand\ee{\end{equation}}
\newcommand\fa{\begin{eqnarray}}
\newcommand\ffa{\end{eqnarray}}
\newcommand{\N}{{\mathbb N}}
\newcommand{\Z}{{\mathbb Z}}
\newcommand{\case}[2]{{\textstyle\frac{#1}{#2}}}
\newcommand{\sgn}{\mathop{\rm sgn}}
\newcommand{\lP}{l_{\rm P}}

\begin{document}

{\renewcommand{\thefootnote}{\fnsymbol{footnote}}
\hfill  CGPG--02/7--1\\
\medskip
\hfill gr--qc/0207038\\
\medskip
\begin{center}
{\LARGE  Isotropic Loop Quantum Cosmology with Matter}\\
\vspace{1.5em}
Martin Bojowald$^a$\footnote{e-mail address: {\tt
bojowald@gravity.psu.edu}} and
Franz Hinterleitner$^{a,b}$\footnote{ e-mail address: {\tt franz@physics.muni.cz}}\\
\vspace{0.5em}
$^a$Center for Gravitational Physics and Geometry,
The Pennsylvania State
University,\\
104 Davey Lab, University Park, PA 16802, USA\\[2mm]
$^b$Department of Theoretical
Physics and Astrophysics,
Masaryk University,\\
Kotl\'{a}\v{r}sk\'{a} 2, 611 37 Brno, Czech Republic\\
\vspace{1.5em}
\end{center}
}

\setcounter{footnote}{0}

\begin{abstract}
A free massless scalar field is coupled to homogeneous and isotropic
loop quantum cosmology. The coupled model is investigated in the
vicinity of the classical singularity, where discreteness is essential
and where the quantum model is non-singular, as well as in the regime
of large volumes, where it displays the expected semiclassical
features. The particular matter content (massless, free scalar) is
chosen to illustrate how the discrete structure regulates pathological
behavior caused by kinetic terms of matter Hamiltonians (which in
standard quantum cosmology lead to wave functions with an infinite
number of oscillations near the classical singularity). Due to this
modification of the small volume behavior the dynamical initial
conditions of loop quantum cosmology are seen to provide a meaningful
generalization of DeWitt's initial condition.

\end{abstract}

\section{Introduction}
\label{sec:1}

Since the early days of canonical quantum gravity \cite{DW,Mi}
isotropic cosmological models have been popular test objects. Due to
the symmetries their number of degrees of freedom is finite and small
so that they become accessible to quantum mechanical methods. In the
geometrodynamical approach quantum states are considered to be
continuous functions of one or more variables, typically metric or
exterior curvature quantities of space-like slices, and of matter
variables if the system is coupled to matter.

In the classical limit at large volume such models are well behaved
and reproduce their classical counterpart.  The problem of the
classical singularity at zero volume, however, is not cured, which is
related to the fact that spectra of geometric operators remain
continuous in standard quantum cosmology. (Nevertheless, at the level
of {\em expectation values\/} the expansion/contraction velocity, for
example, may remain finite and discrete features may emerge
\cite{FH}.) On the other hand, loop quantum cosmology \cite{IL}
inherits from quantum geometry a discrete structure of geometry
\cite{RS,AL} which is most relevant at small scales.  Still, at large
volume standard quantum cosmology emerges as an approximation
\cite{SC} which is very good in the semiclassical regime but not
applicable near the classical singularity, where quantum effects of
gravity become dominant and the discreteness of geometry at the scale
of the Planck length becomes essential. The discreteness leads to a
resolution of the singularity problem through the following mechanism:
Already at the kinematical level, there is an indication for a natural
curvature cut-off since the classically diverging quantity $a^{-1}$
has a finite quantization with an upper bound of the size of the
inverse Planck length \cite{SF}. Furthermore, an investigation of the
dynamical evolution confirms that the classical singularity does not
appear as a boundary but instead allows a well-defined evolution
through it \cite{AS}.

In the context of the present paper we are mostly interested in
another consequence of loop quantum cosmology, namely that it predicts
dynamical initial conditions which are derived from the evolution
equation \cite{DI}. The issue of initial conditions in quantum
cosmology has been widely discussed \cite{DW,HH,VI}, in part as an
attempt to deal with the singularity. DeWitt's initial condition,
which is closely related to the outcome of the dynamical initial
conditions, requires the wave function to vanish at the classical
singularity (which, however, with continuous geometrical spectra does
not solve the singularity problem since one can still get arbitrarily
close to vanishing volume \cite{BI}). On the other hand, it is well
known that DeWitt's condition is not applicable in general because in
most cases it would predict an identically vanishing wave function. A
particularly thorny issue, also for other boundary proposals, is the
fact that solutions to the Wheeler--DeWitt equation often exhibit an
infinite number of oscillations between vanishing scale factor and any
finite value. In such a case, the limit of the wave function for
$a\to0$ is not always well-defined and one cannot even impose initial
conditions at $a=0$. The origin of the infinite number of oscillations
is the kinetic term in a matter Hamiltonian which is proportional to
$a^{-3}$. At small $a$, the Wheeler--DeWitt equation requires the
derivative of the wave function to be proportional to the square root
of the kinetic term, which diverges for $a\to0$. Usually, one tries to
avoid this problem by choosing a wave function which is independent of
the matter field at small $a$ so that the matter momentum vanishes
and the kinetic term is identically zero. However, for larger volume
the wave function {\em must\/} depend on the matter field
non-trivially which forces a dependence at small volume, too. Even
if the dependence is only weak, the diverging kinetic term will
eventually dominate when $a$ is small enough. In the present paper we
investigate if there is a more natural way to deal with this problem
from the point of view of loop quantum cosmology.

Since this problem is caused by the kinetic term independently of the
potential, we will analyze it in the most simple setting, which is
isotropic loop quantum cosmology coupled to a free massless scalar
field. In this way, the total quantum Hamiltonian acts on the state
function containing one degree of freedom of geometry and one of
matter. The formulation is based on \cite{IL,SF} where a discrete
orthonormal basis of states of homogeneous and isotropic quantum
geometry has been established which provides eigenstates of the volume
operator.

This paper is organized as follows. In section 2 the quantum
Hamiltonian constraint equation with a massless scalar field is
formulated, leading to a difference equation which is solved in
section 3. In section 4 the continuum and semiclassical limit are
considered and possible physical consequences are discussed. Section 5
deals with a comparison of different proposals for initial conditions.

\section{The quantum Hamiltonian constraint}
\label{sec:2}

We begin with the Lagrange density of a massless scalar field $\phi$
on a spacelike slice of a canonically decomposed 4-dimensional space
time manifold,
\[
  {\cal L}_{\phi}=\case{1}{2}\left[\dot{\phi}^2-(\nabla\phi)^2\right], 
\]
where $\nabla$ denotes the spatial derivative on the slice and the dot
the derivative w.\,r. to the time coordinate. In our case, the
spatially flat Friedmann model with the metric
\[
 {\rm d}s^2=-{\rm d}t^2+a^2(t)\left[{\rm d}x^2+{\rm d}y^2+{\rm
d}z^2\right], 
\]
the total matter Lagrangian becomes
\[
 L_{\phi}=\int{\rm d}^3x\,\sqrt{-g}\,{\cal L}= \case{1}{2}\int{\rm
 d}^3x\,a^3\,\dot{\phi}^2\,,
\]
where, in order not to disturb homogeneity, $\phi$ is assumed to be
spatially constant so that the gradient term vanishes identically. The
integral over the coordinate depending term, i.\,e. the total volume
divided by $a^3$, will be set equal to one which can be achieved by an
appropriate compactification. For the Hamiltonian of the field we
obtain \be H_{\phi}= \case{1}{2}\,p_\phi^2\, a^{-3}\ee where
$p_{\phi}=a^3\dot{\phi}$ is conjugate to $\phi$.

Together with the gravitational part of the Hamiltonian constraint we
obtain in the isotropic case \be \label{Hamclass}
H=-6\gamma^{-2}\kappa^{-1}c^2\sqrt{|p|}+H_{\phi}=
-6\gamma^{-2}\kappa^{-1}c^2\sqrt{|p|}+\case{1}{2}
|p|^{-\frac{3}{2}}p_{\phi}^2=0 \ee where the canonical gravitational
degrees of freedom are the isotropic connection component $c$ and
(densitized) triad component $p$ which fulfill
$\{c,p\}=\frac{1}{3}\gamma\kappa$. Here, $\kappa=8\pi G$ is the
gravitational constant and $\gamma$ the Barbero--Immirzi parameter
\cite{Im} whose value does not affect the classical behavior (but it
is important for the quantum theory where it controls the continuum
limit). Because $p$ is a triad component which has two possible
orientations, it can take both positive and negative values. This is
also true for the scale factor $a$ which is the isotropic co-triad
component and related to $p$ by $p=\sgn(a) a^2$. {\em From now on, however,
it will be sufficient to consider only positive $p$ and $a$\/} (though we
keep absolute value signs in some formulae for the sake of
generality). The connection component $c$ is related to the extrinsic
curvature and therefore to the time derivative of $a$ by
$c=\frac{1}{2}\dot{a}$.

In the absence of a potential, the matter momentum $p_{\phi}=\omega$
is a constant in time and the solution \be\label{cofp}
c=\case{1}{2}\gamma\sqrt{\kappa/3}\,\omega/p \ee of (\ref{Hamclass})
only depends on $p$. To understand the classical evolution in a
coordinate time $t$ (with lapse function $N=1$) we compute
\[
 \dot{p}=\{p,H\}=4\gamma^{-1}c\sqrt{p}= 2\sqrt{\kappa/3}\,\omega
 p^{-\frac{1}{2}}
\]
yielding $p(t)=a(t)^2=(\sqrt{3\kappa}\,\omega(t-t_0))^{\frac{2}{3}}$,
i.e.\ an eternally expanding universe.

Now we quantize the field canonically by assuming a wave function
$\chi(\phi)$ and the canonical momentum acting as a derivative
operator on it, \be \hat{p}_\phi:=-i\hbar\frac{\rm d}{{\rm d}\phi},\ee
so that the Hamiltonian for spatially constant fields becomes \be
\label{Hmattstand} \tilde{H}_\phi=-\case{1}{2}\hbar^2
|a|^{-3}\,\frac{{\rm d}^2}{{\rm d}\phi^2}\ee (it is denoted
$\tilde{H}_{\phi}$ since we will later introduce another operator
$\hat{H}_{\phi}$ in which also $a$ is quantized).  

For a quantization in the complete system of gravity and matter we
need a quantization of the inverse power $|a|^{-3}$ which diverges
classically close to the singularity. In standard quantum cosmology
this would simply be quantized to a multiplication operator acting on
wave functions depending on $a$, which does not cure the
divergence. Loop quantum cosmology, on the other hand, can easily deal
with this problem: while the volume operator has zero eigenvalues and
so no well-defined inverse, there are well-defined quantizations of
inverse powers of the scale factor \cite{SF}. This inverse power of
the scale factor is essential for the issues studied in the present
paper. The details of the quantization of the matter field is irrelevant;
one can also use quantization techniques inspired from loop quantum
gravity (see e.g.\ \cite{MQ}).

To specify the gravitational part of the wave function in loop quantum
cosmology we start in the connection representation where an
orthonormal basis is given by \cite{IL} \be \langle
c|n\rangle=\frac{\exp(\frac{1}{2}inc)}{\sqrt{2}\sin\frac{1}{2}c}\quad,
\quad n\in\Z\,.\ee These states are eigenstates of the volume operator
\be\label{Vol} \hat{V}|n\rangle=\left(\case{1}{6}\gamma
\lP^2\right)^\frac{3}{2}\sqrt{(|n|-1)|n|(|n|+1)}\;|n\rangle =:
V_{\frac{1}{2}(|n|-1)}|n\rangle\ee and of the inverse scale factor
operator (we only use the diagonal part of the operator $\hat{m}_{IJ}$
derived in \cite{SC}) \be \label{invsc} \widehat{|a|^{-1}}|n\rangle=
16(\gamma\lP^2)^{-2}\left(\sqrt{V_{\frac{1}{2}|n|}}-
\sqrt{V_{\frac{1}{2}|n|-1}}\right)^2|n\rangle \ee with
$\lP=\sqrt{\kappa\hbar}$ being the Planck length (in (\ref{invsc})
$V_{-1}$ is understood to be zero). As discussed in \cite{QA}, the
quantization of the inverse scale factor is affected by quantization
ambiguities. The effect of different choices will be discussed later;
they do not lead to substantial changes in most of the following
calculations and results.

The action of the gravitational Hamiltonian on the basis states
$|n\rangle$ is \cite{IL,TT} \be \label{HE}
\label{Hgrav} \hat{H}_{\rm grav}|n\rangle=3\gamma^{-2}\left(\kappa\gamma
\lP^2\right)^{-1}{\rm
sgn}(n)\left(V_{\frac{1}{2}|n|}-V_{\frac{1}{2}|n|-1}\right)
\left(|n+4\rangle-2|n\rangle +|n-4\rangle\right).\ee

Now we write the states of the coupled system matter plus gravity as a
superposition of geometry eigenstates \be
|s\rangle=\sum_{n=-\infty}^{\infty}s_n(\phi)|n\rangle\ee with
$\phi$-dependent coefficients $s_n(\phi)$ which represent the state in
the triad representation. We use (\ref{invsc}) to quantize the inverse
volume in the matter Hamiltonian (\ref{Hmattstand}). Since the
resulting operator $\hat{H}_\phi$ is diagonal in the basis states
$|n\rangle$ we can define \be \label{Hmattercomp}
\hat{H}_\phi|n\rangle\otimes|\phi\rangle=:|n\rangle\otimes\hat{H}_\phi(n)
|\phi\rangle\ee for each $|n\rangle$ and arbitrary $|\phi\rangle$. For
the massless field we only have to insert the eigenvalue of
$\widehat{|a|^{-1}}$ in a state $|n\rangle$, so we obtain \be
\label{Hmatter} \hat{H}_\phi(n)=-\case{1}{2}\hbar^216^3 (\gamma\lP^2)^{-6}
\left(\sqrt{V_{\frac{1}{2}|n|}}-
\sqrt{V_{\frac{1}{2}|n|-1}}\right)^6\; \frac{{\rm d}^2}{{\rm
d}\phi^2}.\ee

Finally, a state $|s\rangle$ is annihilated by the total Hamiltonian,
the sum of $\hat{H}_{\rm grav}$ in (\ref{Hgrav}) and
$\hat{H}_{\phi}$ in (\ref{Hmattercomp}) with (\ref{Hmatter}), if $s_n$
fulfills (we absorb the sign of $n$ appearing in (\ref{HE}) into the
wave function)
\begin{eqnarray}\label{H}
&&\left(V_{\frac{1}{2}|n+4|}- V_{\frac{1}{2}|n+4|-1}\right)\,
s_{n+4}(\phi)-2\,
\left(V_{\frac{1}{2}|n|}-V_{\frac{1}{2}|n|-1}\right)\,s_n(\phi)\\
&&+\left(V_{\frac{1}{2}|n-4|}- V_{\frac{1}{2}|n-4|-1}\right)\,
s_{n-4}(\phi)=\alpha\hbar^2 \left(\sqrt{V_{\frac{1}{2}|n|}}-
\sqrt{V_{\frac{1}{2}|n|-1}}\right)^6 \frac{{\rm d}^2}{{\rm
d}\phi^2}\,s_n(\phi)\nonumber\end{eqnarray} with
\be\alpha:=\case{2048}{3}\kappa\gamma^2 (\gamma\lP^2)^{-5}.\ee

This is a difference equation of order 8 for the wave function
$s_n(\phi)$ in the internal time $n$. A priori, there could be a
problem at zero volume since some of the coefficients of the difference
equation can vanish due to $V_{-1}=V_{-\frac{1}{2}}=V_0=0$. In fact,
the coefficient of $s_0(\phi)$ --- and only this one --- always
vanishes (note that, unlike in a classical equation,
$\widehat{|a|^{-1}}$ annihilates the zero-volume state which
represents the classical singularity). Thus, there is always a
solution $s_n(\phi)=s_0(\phi)\delta_{n0}$ which is orthogonal to all
other solutions and need not be taken further into account.  The only
free function then is $s_n(\phi)$ for $n=4$, from which one can
compute $s_8$, and so on, yielding all coefficients $s_{4m}(\phi)$ as
functions of $s_4(\phi)$. The intermediate values $s_{4m+i}(\phi)$ for
$i=1,2,3$ are essentially fixed by the principal series $s_{4m}(\phi)$
by requiring that the wave function does not vary strongly on small
scales (i.e.\ that it is pre-classical \cite{DI}).

The simplest choice for the free function $s_4(\phi)$, besides a
constant, is an eigenfunction of the matter Hamiltonian
$\hat{H}_\phi(4)$ for $n=4$, \be \label{init}
s_4(\phi)=\chi(\phi):= e^{i\omega\phi/\hbar}\ee where $\omega$ is the
eigenvalue of $\hat{p}_{\phi}$. (A spatially constant
massless field on Minkowski space with a non-vanishing energy
eigenvalue would not be possible, but here geometry acts as a
potential.) Because $\chi(\phi)$ is an eigenfunction for all
$\hat{H}_\phi(n)$, the $\phi$-dependence of all the $s_{4m}$ is the
same. This will simplify the analysis since we can compute the wave
function from an ordinary difference equation rather than a
difference-differential equation; the combination
$s_n(\phi)e^{-i\omega\phi/\hbar}$ will be $\phi$-independent.

We do not intend to justify the initial condition (\ref{init}) with
any given $\omega$ by a physical argument; rather, we choose it to
simplify the subsequent calculations while retaining qualitative
aspects as to the behavior of the wave function $s_n(\phi)$ close to
the classical singularity in the presence of matter. The advantage of
an initial condition like (\ref{init}) is that the gravitational and
the matter degree of freedom separate such that the $n$-behavior of
the wave function is still given by an ordinary difference
equation. If a more complicated function than (\ref{init}) is
required, it can always be constructed as a suitable superposition of
our solutions with different $\omega$.

\section{Solutions of the Hamiltonian constraint}
\label{sec:3} 

Having made the above observations about $s_0$ and $s_4$, we begin with
inserting $n=4$ into (\ref{H}) and obtain $s_8$ in terms of
$s_4=\chi(\phi)$, 
\[
 s_8(\phi)=\frac{2(V_2-V_1)-\alpha\omega^2(\sqrt{V_2}-
 \sqrt{V_1})^6}{V_4-V_3}\,\chi(\phi)
\]
and, in consequence, all
\begin{eqnarray}\label{s4m}
 s_{4m}(\phi) &=&\frac{2(V_{2m-2}-V_{2m-3})-\alpha\omega^2(\sqrt{V_{2m-2}}-
 \sqrt{V_{2m-3}})^6}{V_{2m}-V_{2m-1}} s_{4m-4}(\phi)\\
 &&- \frac{V_{2m-4}-V_{2m-5}}{V_{2m}-V_{2m-1}} s_{4m-8}(\phi)
 \qquad\mbox{for } m\in\N\,. \nonumber
\end{eqnarray}

For an explicit calculation of the coefficients $s_{4m}$ it is
convenient to introduce some shorthand notations. We define
\[
 D_n:=V_n-V_{n-1}\hspace{2cm}\mbox{and}\hspace{2cm}
 W_n:=\left(\sqrt{V_n}-\sqrt{V_{n-1}}\right)^6\,,
\]
then we split off the $\phi$-dependence by defining $\phi$-independent
coefficients $r_m$ by $r_0=0$, $r_1=1$ and
\[
 s_{4m}(\phi)=:\frac{r_{m}}{D_{2m}D_{2m-2}\ldots D_4}\,\chi(\phi)
 \qquad\mbox{for }m\geq 2\,.
\]
Inserting this into (\ref{s4m}) yields a recurrence relation for the
$r_{m}$: \be \label{rmrec}
r_{m}=\left(2D_{2m-2}-\alpha\omega^2W_{2m-2}\right)
r_{m-1}-D_{2m-2}D_{2m-4}\cdot r_{m-2}.\ee An explicit solution can be
found after calculating a few $r_{m}$, and confirmed by induction:
\begin{equation} \label{rmexpl} r_{m}=
\sum_{k=0}^{m-1}(-\alpha\omega^2)^k\!\!\!
\sum_{\{j_1,\ldots,j_k\}}j_1(j_2-j_1)\cdots(j_k-j_{k-1})
(m-j_k)W_{2j_1}\cdots W_{2j_k}D_{2j_{k+1}}\cdots
D_{2j_{m-1}}.
\end{equation} 
This formula has the following meaning: $j_1,\ldots,j_k$ are integers
between 1 and $m-1$, according to the $\left(\begin{array}{c} m-1 \\ k
\end{array}\right)$ possibilities to choose $k$ numbers out of
$\{1,2,\ldots,m-1\}$, arranged in increasing order. The numbers
$j_{k+1},\ldots,j_{m-1}$ appearing as labels of $D_n$ are given by the
remaining values in $\{1,\ldots,m-1\}\backslash\{j_1,\ldots,j_k\}$ in
an arbitrary ordering.  The case $k=0$ is to be understood in the way
that from the factors containing $j$ there remains only the last one,
$m-j_0$, with $j_0$ defined to be zero, so that this contribution is
$mD_2\cdots D_{2(m-1)}$.

For the purpose of numerically calculating the coefficients in terms
of a given initial one the recurrence relation (\ref{rmrec}) may be
more convenient than the solution (\ref{rmexpl}).

Inserting $n=0$ into (\ref{H}) leads to $s_{-4}(\phi)=s_4(\phi)$. Further
on, (\ref{H}) is symmetric in the sense that the relations between
$s_{-4}$, $s_{-8}$, $s_{-12}$,\ldots\ are the same as the respective
relations for $s_4$, $s_8$, $s_{12}$,\ldots\ and so the series
$s_{4m}$ is symmetric in $m$. By the argument of pre-classicality
\cite{SC} this symmetry applies to the remaining series
$s_{4m+i}$ as well. With the physical interpretation of $n$ as an
internal time we have obtained time symmetry of our wave function
with respect to the classical singularity.

\section{Continuum and semiclassical limit}
\label{sec:4}

For large $n$, when $V_n$ becomes very large in comparison with
$\lP^3$, the discreteness should only lead to small corrections,
i.\,e. the discrete time evolution from $n$ to $n+4$ should become
well approximated by a continuous evolution.  Under this assumption
the difference equation should be approximated with high accuracy by a
differential equation for a continuous wave function
$\psi(p,\phi)=s_{n(p)}(\phi)$, where $n(p)=6p/\gamma\lP^2$ can be
derived from the volume eigenvalues (\ref{Vol}). As discussed in
\cite{DI}, among the solutions of the difference equation there is
always one which is slowly varying at small scales for large $n$ and
so justifies the assumption of almost continuous behavior.

\subsection{Large volume behavior}

To approximate (\ref{H}) by a differential equation it is convenient
first to make a slight change of variables. We define the new variable
(working only with positive $n>0$) \be t_n(\phi):=(\gamma\lP^2)^{-1}
\left(V_\frac{n}{2}-V_{\frac{n}{2}-1}\right) \,s_n(\phi)\ee which for
large $n$ is approximated by $t_n(\phi)\sim
\frac{1}{2}\sqrt{p(n)}\psi(p(n),\phi)=: \tilde{\psi}(p(n),\phi)$.  In
terms of $t_n(\phi)$ the evolution equation becomes \be \label{tdiff}
t_{n+4}(\phi)-2t_n(\phi)+t_{n-4}(\phi)=
\alpha\hbar^2\frac{\left(\sqrt{V_\frac{n}{2}}-
\sqrt{V_{\frac{n}{2}-1}}\right)^6}
{V_\frac{n}{2}-V_{\frac{n}{2}-1}}\,\frac{{\rm d}^2}{{\rm
d}\phi^2}\,t_n(\phi).\ee Furthermore, we need the asymptotic
approximations
\begin{eqnarray*}
  V_\frac{n}{2}-V_{\frac{n}{2}-1} &\sim& 24
 ^{-\frac{1}{2}}(\gamma\lP^2)^{\frac{3}{2}} n^\frac{1}{2}=
 \case{1}{2}\gamma\lP^2\sqrt{p(n)}\\
 \left(\sqrt{V_\frac{n}{2}}-\sqrt{V_{\frac{n}{2}-1}}\right)^6
 &\sim&\left(\case{3}{128}\right)^\frac{3}{2}(\gamma\lP^2)^{\frac{9}{2}}
 n^{-\frac{3}{2}}=(\case{1}{4}\gamma\lP^2)^{6} p(n)^{-\frac{3}{2}}\,.
\end{eqnarray*}
For the differences operator on the left hand side of (\ref{tdiff}) we
obtain in the continuum limit $16\,\partial^2/\partial
n^2=\frac{4}{9}(\gamma\lP^2)^2\partial^2/\partial p^2$, acting on the
function $\tilde{\psi}(p,\phi)$, so the asymptotic equation becomes
\be
\label{asdiff}
\left[\frac{4\lP^4}{3\kappa\hbar^2}\,p^2\frac{\partial^2}{\partial
p^2}-\frac{\partial^2}{\partial\phi^2}\right]\tilde{\psi}(p,\phi)=0.\ee
Note that a standard quantization in geometrodynamics would have
ordering ambiguities in the gravitational part of the Wheeler--DeWitt
equation, whereas the derivation here leads to a unique ordering in
loop quantum cosmology \cite{IL}.

The variables in this partial differential equation can be separated
by a product ansatz,
\[ 
 \tilde{\psi}(p,\phi)=N(p)\chi(\phi)\,,
\]
so that we obtain
\[
 \frac{4\lP^4}{3
 \kappa\hbar^2}\,p^2\frac{N''}{N}=\frac{\chi''}{\chi}=
 -\omega^2/\hbar^2= \mbox{const.}
\]
We assume $\omega^2>0$, because the matter Hamiltonian is expected to
have a positive spectrum. With this assumption the solution $\chi$
coincides with the function used for $s_4$ in (\ref{init}), denoted by
the same letter. The ordinary differential equation for $N(p)$ has the
solutions $N(p)=p^\lambda$ with \be \label{lambda}
\lambda=\case{1}{2}\pm
\sqrt{\case{1}{4}-\case{3}{4}\,\kappa\omega^2\lP^{-4}}.\ee The
asymptotic differential equation (\ref{asdiff}) applies also for
$t_{4m+i}$ for $i=1,2,3$, so, in order to have a smooth function for
large $n$ we assume the $\phi$-dependence of the remaining three free
coefficients $s_1$, $s_2$ and $s_3$ to be the same as that of $s_4$,
given by the function $\chi(\phi)$, endowed with constant factors
which could be determined by a numerical analysis.

\begin{figure}[ht]
\begin{center}
 \includegraphics[width=14.5cm,height=11cm,keepaspectratio]{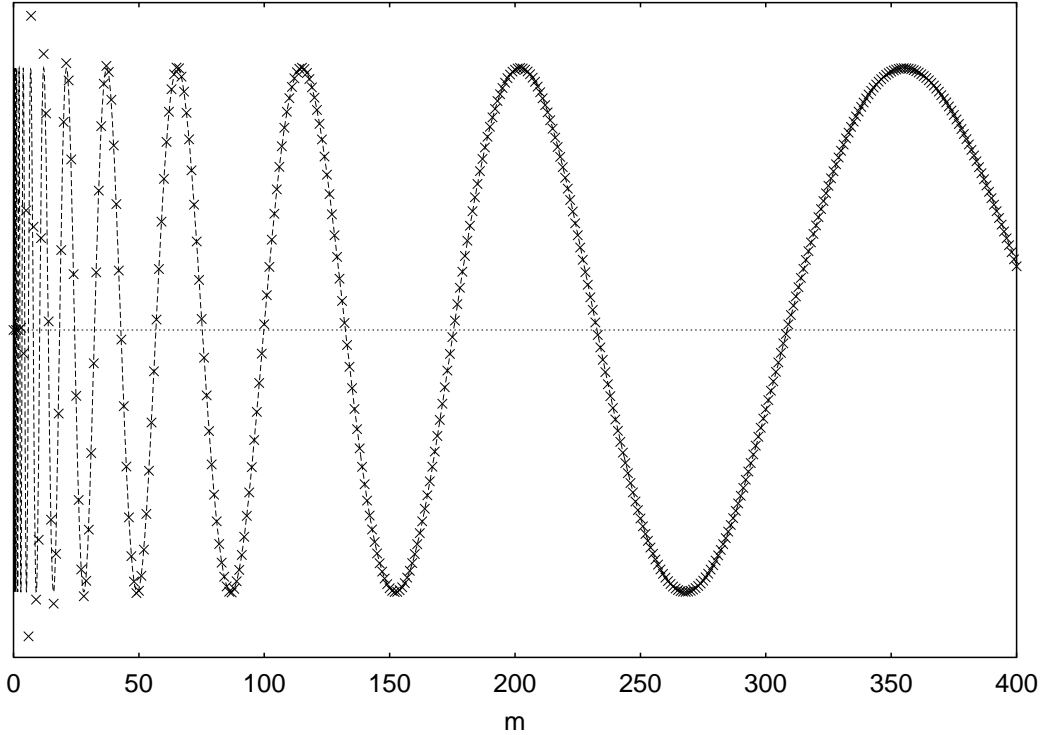}
\end{center}
\caption{The $\phi$-independent part
$s_{4m}(\phi)e^{-i\omega\phi/\hbar}$ of the discrete wave function
($\times$) compared to a continuous solution to the Wheeler--DeWitt
equation (dashed line) for $\omega^2=500\omega_c^2$. Strong deviations
occur only at very small $n=4m$ (see Fig.~\ref{Mattersmall}) and will
be discussed later.}
\label{Matter}
\end{figure}

The last result (\ref{lambda}) deserves some discussion, because it
provides a threshold for a qualitatively different behavior of the
solution $N(p)$, according to the sign of the expression under the
square root. If $\omega$ is smaller than a critical value, determined
by \be \omega_c^2=\case{1}{3}\lP^4/\kappa=
\case{1}{3}\kappa\hbar^2\,, \ee the exponent $\lambda$ is real
and for $\omega$ going to zero one solution for $N(p)$ approaches $p$,
the asymptotic function for the vacuum calculated in \cite{IL}, the
other one approaches a constant. For $\omega$ larger than $\omega_c$
\be \label{Np} N(p)=p^\frac{1}{2}p^{\pm i\Omega}=p^\frac{1}{2}\,e^{\pm
i\Omega\log p}\ee with \be \label{Om}
\Omega=\case{1}{2}\sqrt{3\kappa\omega^2\lP^{-4}-1}=
\case{1}{2}\hbar^{-1}\sqrt{3\omega^2/\kappa-\hbar^2}.\ee

\begin{figure}[ht]
\begin{center}
 \includegraphics[width=14.5cm,height=11cm,keepaspectratio]{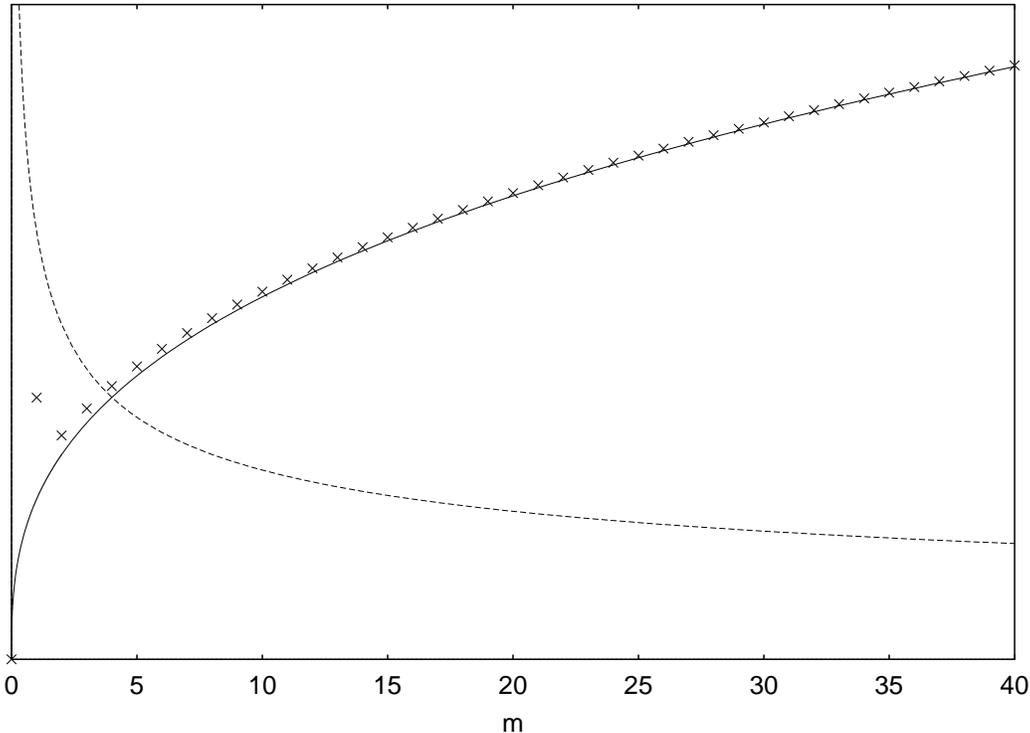}
\end{center}
\caption{A subcritical ($\omega^2=\frac{1}{2}\omega_c^2$) wave
function $s_{4m}(\phi)e^{-i\omega\phi/\hbar}$ compared with the
continuous DeWitt wave function (solid line) and a continuous wave
function which does not fulfill DeWitt's initial condition (dashed
line; any such solution diverges at $a=0$).}
\label{Mattersub}
\end{figure}

Reconstructing $\psi(p,\phi)$ from $\tilde{\psi}(p,\phi)$ we finally
obtain \be
\label{snphi} \psi(p,\phi)=2 e^{\pm i\Omega\log p}
e^{i\omega\phi/\hbar}.\ee Formally this solution is a quantum
mechanical wave function of a particle moving from small values of $p$
to larger ones with decreasing speed, whereas solutions with real
powers of $p$, coming from undercritical values of $\omega$, are
lacking such a dynamical interpretation. An example of an oscillating
solution can be seen in Fig.~\ref{Matter} for
$\omega^2=500\omega_c^2$, a non-oscillating one with
$\omega^2=\frac{1}{2}\omega_c^2$ in Fig.~\ref{Mattersub}.

Note that the presence of a critical frequency $\omega_c$ depends on
the ordering of the Wheeler--DeWitt equation; choosing, e.g., the
ordering $p\partial/\partial p(p\partial/\partial p)$ instead of
$p^2\partial^2/\partial p^2$ would give oscillating solutions for all
non-zero $\omega$. Therefore, standard quantum cosmology cannot imply
such a behavior reliably. Loop quantum cosmology, on the other hand,
has a distinguished ordering and thus predicts the existence of a
critical frequency for the system studied here.

Before interpreting the physical significance of $\omega_c$ we note
that the possibility of a non-oscillating wave function for
$\omega^2<\omega_c^2$ is not in contradiction with semiclassical
behavior at large volume.  The standard semiclassical analysis
requires oscillating (WKB) solutions which in our case can be derived
from (\ref{cofp}) after replacing $c$ with
$\frac{1}{3}\gamma\kappa\partial S/\partial p$ and are given by
$N_{WKB}(p)=\exp(iS/\hbar)= \exp(i\Omega_{WKB}\log p)$ with
$\Omega_{WKB}=\frac{1}{2}\sqrt{3\kappa}\,\omega/\lP^2$. For
$\omega^2\gg\omega_c^2$ the frequency $\Omega$ in fact reduces to
$\Omega_{WKB}$. The non-oscillating solutions for
$\omega^2<\omega_c^2$ cannot be interpreted in this way, but this
should not be expected since in the classical limit
$\omega_c=\frac{1}{3}\kappa\hbar^2$ vanishes and so any
non-zero frequency will be larger than the critical one.

\subsection{A critical energy}

We have seen that loop quantum cosmology predicts the presence of a
critical frequency for a dynamical evolution in the particular model
studied here. While this prediction is reliable (in contrast to
standard quantum cosmology where it may or may not exist, depending on
the factor ordering of the constraint), we will see that the precise
value depends on quantization ambiguities of a different kind.

For an interpretation of $\omega_c$ we first compute the associated
($1/a$-dependent) eigenvalues of the matter Hamiltonian, \be
E(a)=\case{1}{2}|a|^{-3}\,\omega^2.\ee In standard quantum cosmology,
this expression is unbounded from above and does not have a
distinguished value which would be suitable for an interpretation. In
loop quantum cosmology, on the other hand, we do have --- for a given
$\omega$ --- an upper bound for $E$ which can serve as a natural value
for an interpretation. Taking for $1/|a|$ the maximal eigenvalue of
the inverse scale factor operator, occurring at $n=2$, we get a
relation between the maximal field energy concentrated in
$V_{1/2}=\case{1}{6}(\gamma\lP^2)^{\frac{3}{2}}$ and $\omega^2$:
\[
 E_{\rm max}=\frac{2^{11}}{3^3(\gamma\lP^2)^{\frac{3}{2}}}\,
 \omega^2\,.
\]
By inserting $\omega^2$ into (\ref{Om}) the distinction between
oscillating and non-oscillating behavior can be expressed in terms of
this initial energy, \be
\Omega=\frac{9\sqrt{\pi}}{32}\gamma^{\frac{3}{4}} \sqrt{\frac{E_{\rm
max}}{E_{\rm P}}-\frac{256}{81\pi}\gamma^{-\frac{3}{2}}},\ee with
$E_{\rm P}$ denoting the Planck energy $8\pi\hbar/\lP=\lP/G$.  This
expresses $\Omega$ in terms of the ratio $E_{\rm max}/E_{\rm P}$,
whose critical value $\frac{256}{81\pi}\gamma^{-\frac{3}{2}}\approx
1.006$ is close to one for $\gamma$ equal to one, but has the larger
value $\approx 21.5$ for the value $\gamma=\log 2/\pi\sqrt{3}\approx
0.13$ which has been computed by comparing the black hole entropy
resulting from a counting within quantum geometry with the
semiclassical Bekenstein--Hawking result \cite{BH1,BH2}. Only if
$E_{\rm max}/E_{\rm P}$ is larger than the critical value can a
dynamical evolution of the classical Friedmann model set in.

This observation implies that we need a maximal energy in a Planck
volume which {\em exceeds\/} the Planck energy, being apparently in
conflict with the heuristic but widespread expectation that there can
be at most an energy amount of $E_{\rm P}$ in a Planck volume. (Often,
the holographic principle \cite{HP} is used to arrive at this
conclusion. In the present context, there is no direct contradiction
of results obtained within the formalism used here.)  Usually,
inflation is invoked to explain how the huge amount of energy in the
present universe can have originated from a Planck scale
universe. Here we can see that the relation between the maximal energy
and inflation is even more intimate in loop quantum cosmology: One can
reduce the critical value for $E_{\rm max}/E_{\rm P}$ by using a
different quantization for the inverse scale factor instead of
(\ref{invsc}) parameterized by an ambiguity parameter $j$ (a non-zero
half-integer, see \cite{QA}) such that
\begin{equation}\label{aj}
 \widehat{|a|_j^{-1}}|n\rangle= 144 (j(j+1)(2j+1))^{-2}(\gamma\lP^2)^{-2}
 \left(\sum_{k=-j}^j k\sqrt{V_{\frac{1}{2}(|n+2k|-1)}}\right)^2\,.
\end{equation}
As derived in \cite{QA}, the maximal value of $|a|_j^{-3}$ is then
attained if $n=2j$ and approximately given by $V_j^{-1}$ (the
approximation gets better for large $j$). This gives a maximal energy
\[
 E_{{\rm max},j}\sim\case{1}{2}\omega^2V_j^{-1}\sim
 \case{1}{2}\omega^2(\gamma\lP^2)^{-\frac{3}{2}} (3/j)^{\frac{3}{2}}
\]
and
\[
 \Omega\sim \sqrt{4\pi}3^{-\frac{1}{4}} (\gamma
 j)^{\frac{3}{4}}\sqrt{ \frac{E_{{\rm max},j}}{E_{\rm
 P}}-\frac{\sqrt{3}}{16\pi (\gamma
 j)^{\frac{3}{2}}}}\,.
\]
The critical value $\case{\sqrt{3}}{16\pi}(\gamma
j)^{-\frac{3}{2}}\approx 0.74j^{-\frac{3}{2}}$ for $E_{\rm max}/E_{\rm
P}$ is then suppressed by $j^{-\frac{3}{2}}$ and much smaller than one
for large $j$ (note that this critical value for $j=\frac{1}{2}$ does
not coincide with the one obtained above because the approximation for
the maximum of (\ref{aj}) is bad for very small $j$). Moreover, the
maximal energy is not obtained in the initial Planck volume but in a
volume of size $V_j\sim (\case{1}{3}j\sqrt{\gamma}\lP)^3$. As observed
in \cite{In}, using a quantization with a large value of $j$ also
leads to a prolonged phase of inflation in the very early universe,
displaying the relation between a maximal energy below the Planck
energy and inflation.

\begin{figure}[ht]
\begin{center}
 \includegraphics[width=14.5cm,height=11cm,keepaspectratio]{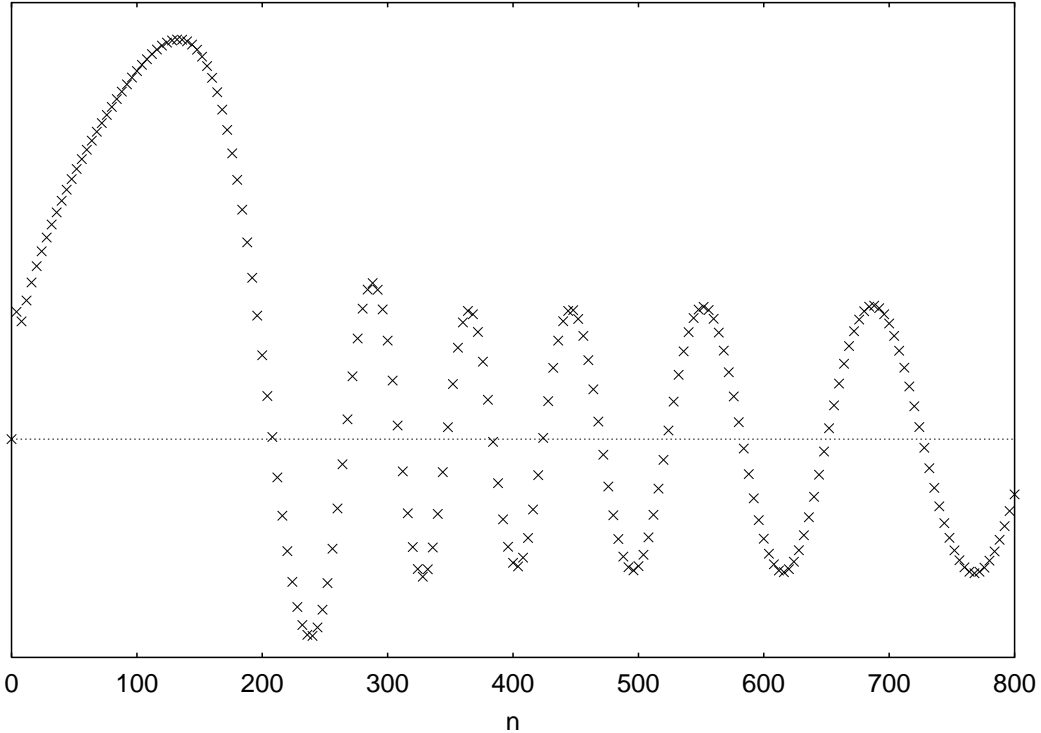}
\end{center}
\caption{A wave function with an extended inflationary phase ($j=200$)
and standard behavior at large volume for $n>2j=400$
($\omega^2=500\omega_c^2$).}
\label{Matterinfl}
\end{figure}

A large value of $j$ modifies the small volume behavior of the wave
function considerably (Fig.~\ref{Matterinfl}). Its oscillation length
and amplitude first decrease with increasing volume, which is
characteristic of an accelerating universe. When the maximal value for
$|a|_j^{-3}$ is reached, standard behavior (\ref{snphi}) sets in with
increasing oscillation length toward larger volume.

\section{Comparison with continuum initial conditions}

\begin{figure}[ht]
\begin{center}
 \includegraphics[width=14.5cm,height=11cm,keepaspectratio]{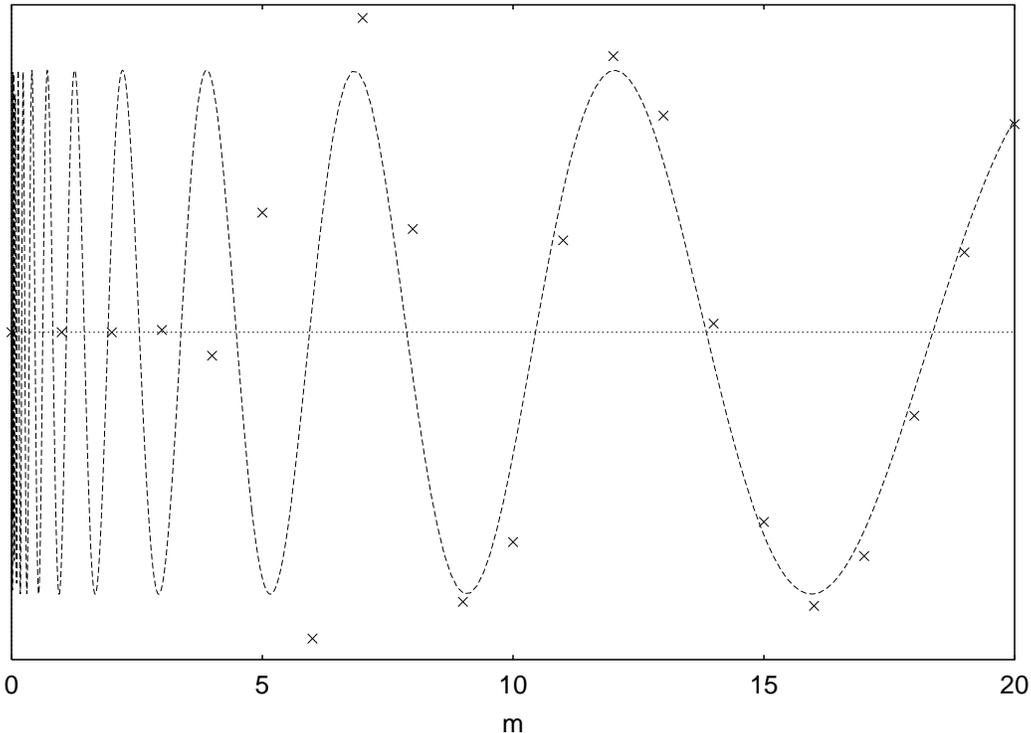}
\end{center}
\caption{The small-$m$ behavior of the $\phi$-independent part
$s_{4m}(\phi)e^{-i\omega\phi/\hbar}$ of the discrete wave function
($\times$) compared to a continuous solution to the Wheeler--DeWitt
equation (dashed line) for $\omega^2=500\omega_c^2$. Whereas the
continuous solution displays an infinite number of oscillations, the
discrete solution decreases toward zero once the oscillation length
is comparable to the Planck scale.}
\label{Mattersmall}
\end{figure}

In loop quantum cosmology the initial conditions for the gravitational
part of the wave function are derived from the evolution equation and
thus fixed \cite{DI}. While $s_0$, the value of the wave function at
the classical singularity, drops out of the evolution equation and so
remains unspecified, the lowest values of $n$ show that the wave
function approaches the value zero for small $n$ (see
Figs.~\ref{Matter} and \ref{Mattersmall}). Interpreted as an initial
value, this is reminiscent of DeWitt's proposal that the continuum
wave function $\psi(a)$ should vanish for $a=0$ \cite{DW}. It is well
known that DeWitt's initial condition can be satisfied non-trivially
only in special systems (e.g., for quantum de Sitter space);
generically it would imply that the wave function vanishes
identically.

An example for a system where DeWitt's condition does not work in
general is the one discussed in the present paper: (\ref{snphi}) is a
continuous wave function which does not have a well-defined limit for
$a\to0$ if $\omega^2>\omega_c^2$ because it oscillates with constant
amplitude and diverging frequency close to the classical singularity
(for $\omega^2<\omega_c^2$ DeWitt's initial condition can be defined
and coincides with the result of the dynamical initial condition; see
Fig.~\ref{Mattersub}). Another example is \cite{PF} where a wave
function of the same kind as (\ref{snphi}) was obtained for stiff
matter and rejected because it does not satisfy a DeWitt boundary
condition in the sense that either the function or its derivative
should be equal to zero at the classical singularity. The same is true
for the continuum approximation of our wave function, but in this case
the problem is cured by discreteness near the singularity.

The oscillations are suppressed once their oscillation length becomes
smaller than the Planck length, and consequently their amplitude
approaches zero. (If there is prolonged inflation with a large value
of $j$, as discussed above, the oscillation length may never reach the
Planck length where this mechanism would be necessary. For a fixed
$j$, however, a regime with small oscillation length will always be
present if $\omega$ is large enough.) This can be seen by following
the analysis of \cite{DI}: If we define
\[
 P(n):=\case{1}{3}\kappa \gamma\lP^2 H_{\phi}(n)(V_{n/2}-V_{n/2-1})^{-1}
\]
which at large volume is approximately $P\sim\case{2}{3}\kappa
H_{\phi}/a$, the discrete evolution equation (\ref{tdiff}) is
\[
 t_{n+4}-(2-\gamma^2P(n))t_n+t_{n-4}=0\,.
\]
In the neighborhood of a fixed $n$ we can assume that $P(n)$ is a
constant and solve the difference equation with an ansatz
$t_n=e^{in\theta}$ which leads to
\[
 e^{4i\theta}-(2-\gamma^2 P)+e^{-4i\theta}=2\cos 4\theta-2+\gamma^2
 P=0\,.
\]

If $P$ is small, the solution $\theta=\frac{1}{4}\arccos(1-\gamma^2
P/2)$ is real and small, implying an oscillating wave function
$t_n=e^{in\theta}$ with long oscillation length
$2\pi/\theta\approx\pi/\gamma\sqrt{P}$ as observed at large
volume. (This conclusion is only valid if the resulting oscillation
length is smaller than the length scale on which $P(n)$ changes
substantially. In particular, it does not apply to subcritical
solutions for which the predicted oscillation length would be too
large.)  As $P$ becomes larger, which will happen for decreasing $a$,
$\theta$ will first increase up to $\theta_{\rm max}=\pi/4$ (for
$P=2\gamma^{-2}$) and then become imaginary. The result is that the
wave function will first oscillate more and more rapidly, as expected
from the continuum approximation, but then enter a branch with
exponential behavior which does not appear in the continuum
formulation. Of the two independent solutions --- exponentially
increasing or decreasing --- only the decreasing one is allowed by the
dynamical initial conditions (Fig.~\ref{Mattersmall}). This is the
mechanism, essentially relying on the discreteness, which allows to
generalize DeWitt's initial condition to systems where the continuum
version does not work. (Note that a similar behavior with exponential
solutions can occur at large volume in the presence of a positive
cosmological constant. In such a case, it would be caused by large
volume rather than large curvature and therefore signals an infrared
problem. In the small volume regime, however, a modified behavior is
perfectly admissible and welcome since the classical description is
expected to break down if curvatures become large.)

The diverging oscillation number close to $a=0$ renders inapplicable
not only DeWitt's proposal but any condition which requires properties
of the continuous wave function at $a=0$, which includes the
``no-boundary'' and the ``tunneling'' proposals. (Usually, this
problem is avoided by setting $\omega=0$ or an analogous condition by
hand, see e.g.\ \cite{KW}. Note that this is a very strong assumption
since even a tiny $\omega$ would eventually yield a large kinetic term
due to the inverse volume.) As already discussed, of the proposed
initial conditions only the discrete formulation of loop quantum
cosmology can deal with the problem of wild oscillations close to the
classical singularity.

\section{Conclusions}

In this paper we considered a free massless scalar field coupled to
loop quantum cosmology as a model for implications of the kinetic term
in a matter Hamiltonian. This term diverges classically at zero volume
and is also problematic in standard quantum cosmology where it leads
to infinitely many oscillations of the wave function close to the
classical singularity. It is caused by the inverse volume in the
kinetic term and is not sensitive to the special form of matter or its
quantization. Therefore, we restricted our attention to the free
massless scalar field quantized in a standard way as usual in the
Wheeler--DeWitt approach. A possible mass or potential term, which is
proportional to the volume, would not change the qualitative behavior
at small enough volume where it would be suppressed. Also higher spin
fields or different quantization techniques applied to the matter
field, e.g.\ inspired by methods for full quantum geometry \cite{MQ},
would not be significant as far as the present paper is concerned. We
also note that we simplified the analysis by using only the Euclidean
part of the gravitational constraint multiplied with $\gamma^{-2}$
instead of the full Lorentzian constraint in (\ref{HE}).  In the flat
case, both expressions agree classically, and at the quantum level the
more complicated Lorentzian constraint does not lead to significant
changes of the qualitative behavior.

The quantization of the gravitational degrees of freedom, on the other
hand, is important since it affects the form of the inverse volume in
a quantization of the kinetic term. While standard quantum cosmology
treats the inverse volume as a multiplication operator which does not
cure its divergence, loop quantum cosmology leads to a quantization
with significant changes at small scales where the discreteness of the
volume is important and its inverse does not diverge. This has already
been seen to imply the absence of cosmological singularities
\cite{AS}, dynamical initial conditions for the wave function of a
universe \cite{DI}, and a new origin of inflation \cite{In}. The main
result of the present paper is that the discreteness at small scales
also leads to a cure of pathologies of the wave function in a standard
quantization, like the infinite number of oscillations caused by a
kinetic matter term.

Also the issue of initial conditions is further elucidated by this
analysis. In \cite{DI} it has been shown that loop quantum cosmology
predicts dynamical initial conditions for the wave function of a
universe which can be derived from the evolution equation and need not
be imposed by hand. In this derivation the structure of the cured
classical singularity plays an important role. In their effect on the
wave function the dynamical initial conditions resemble most closely
DeWitt's initial condition that the wave function should vanish at the
classical singularity. The drawback of DeWitt's approach, namely that
this initial condition cannot be fulfilled non-trivially for most
systems of physical interest, has been seen to be eliminated by
effects of the discreteness of loop quantum cosmology. Therefore, {\em
the dynamical initial conditions of loop quantum cosmology present a
meaningful generalization of DeWitt's initial conditions}.

We also emphasize that loop quantum cosmology leads to more reliable
results than standard quantum cosmology because some quantization
ambiguities like the factor ordering of the gravitational part of the
constraint are fixed. Nevertheless there are ambiguities of a
different kind which can affect physical consequences and therefore in
principle lead to observable effects.  This has played a role in the
discussion of a critical matter energy necessary for an oscillating
wave function. Its value depends on the quantization ambiguities, but
its presence can be concluded from loop quantum cosmology, unlike
standard quantum cosmology where it may or may not occur depending on
the ordering.

\subsection*{Acknowledgements}

We are grateful to A.~Ashtekar, D.~Coule and S.~Major for discussions.
M.~B.\ was supported in part by NSF grant PHY00-90091 and the Eberly
research funds of Penn State. F.~H. would like to acknowledge CGPG
for hospitality at Penn State and the Czech ministry of education for
support, contact no 143100006.

\end{document}